# Continuous Space-Time Crystal State Driven by Nonreciprocal Optical Forces

Venugopal Raskatla[1], Tongjun Liu[1], Jinxiang Li[2], Kevin F. MacDonald[1], and Nikolay I. Zheludev[1,2]

[1]Optoelectronics Research Centre, University of Southampton, Highfield, Southampton, SO17 1BJ, UK

[2]Centre for Disruptive Photonic Technologies, TPI & SPMS, Nanyang Technological University, Singapore, 637371

Continuous time crystals (CTCs) – media with broken continuous time translation symmetry - are an eagerly sought state of matter that spontaneously transition from a time-independent state to one of periodic motion in response to a small perturbation. The state has been realized recently in an array of nanowires decorated with plasmonic metamolecules illuminated with light. Here we show that this as-yet-unexplained CTC state can be understood as arising from a nonreciprocal phase transition induced by nonconservative radiation pressure forces among plasmonic metamolecules: above a certain intensity threshold, light drives the inhomogeneously broadened array of thermally-driven noisy nanowire oscillators to a synchronized coherent space-time crystal state and ergodicity of the system is broken. At the onset of synchronization, this mechanism does not require nonlinearity in the oscillators but depends instead on nonreciprocal forces. As such it is fundamentally different from the regimes of synchronization that depend on nonlinearity.

Over the past decade, the quest for experimentally realized time crystals [1, 2] – a state of matter with spontaneously broken time-translation symmetry – has been vigorously pursued. It is now understood that nature prohibits closed time crystal systems. However, time crystals with discretely broken time-translation symmetry have been demonstrated in trapped ions, atoms, and spins where an external periodic force spontaneously initiates oscillation in the system at the sub-harmonic frequencies [3]. Systems that break continuous time translation symmetry realize the spirit of the original proposal more closely than discrete time crystals and represent a new state of matter. A continuous time crystal (CTC) is a many-body system in which continuous time translation symmetry is spontaneously broken into a periodic motion in response to an arbitrarily weak perturbation. CTC behaviour has been seen first in the form of slow oscillation dynamics in the nonlinear electron-nuclear spin system of a semiconductor at Kelvin temperatures [4, 5]. Recently CTCs have been realized in an optically pumped dissipative Bose-Einstein condensate in an optical cavity [6] and in a strongly interacting Rydberg gas at room temperature [7].

These observations of CTCs in sophisticated quantum atomic and spin systems have demonstrated the feasibility of the state but did not create CTC materials offering practical applications. However, it has recently been shown experimentally that a simpler classical system – an array of dielectric nanowires decorated with plasmonic metamolecules (a metamaterial), can be optically driven into the state of robust time-space persistent coherent oscillations, representing the first demonstration of a continuous *space-time crystal,* a material potentially deployable in electrooptical, timing and sensing applications [8].

We show that the transition to the CTC state belongs to the recently identified class of phase transitions where nonreciprocity of interactions in a many-body system leads to time-dependent phases where spontaneously broken continuous time translation symmetries are dynamically restored to states with discreetly broken time translation symmetries. It has also been theoretically shown and illustrated with simple robotic demonstrations that nonreciprocal interaction may lead to synchronization, flocking and pattern formation that can be described the in the framework of bifurcation theory and non-Hermitian quantum mechanics [9].

In our paper here we demonstrate for the first time that light can be the agent of nonreciprocity in such phase transitions. In a non-Hamiltonian ensemble of oscillators, in the presence of light that pumps energy into the system, nonreciprocal forces can emerge from the radiation pressure induced by the scattered fields. These light-induced forces synchronize the oscillator movements and break the ergodicity of the system in the transition to the CTC state.

Nonreciprocal synchronization relies on the fact that the optical forces between a pair of objects (e.g. particles) interacting in a light field can be nonreciprocal. Here, we refer to light-induced forces between two objects breaking the action-reaction equality, rather than the reciprocity of electromagnetic field propagation introduced through the Lorentz and Feld-Tai lemmas. Optical forces are generally perceived to be conservative and reciprocal, e.g. optical trap forces proportional to the gradient of light intensity. However, light fields can also induce nonconservative forces of radiation pressure from scattered fields, which constantly pump energy into a system [10-13].

The nonconservative component can be significant in nanoscale systems, leading to apparent violations of the action-reaction equality, as illustrated in Fig. 1 for the simple case of two closely spaced plasmonic nanoparticles under continuous plane wave coherent illumination. When the particles are of the same size and aligned with their centre-to-centre axis parallel to the incident wavefront (Fig. 1a), the light-induced forces acting on the particles along that axis are reciprocal ($\boldsymbol{F}_1 = -\boldsymbol{F}_2$). However, if the particles are illuminated with a mutual phase lag (Fig. 1b) or they differ in size/shape (Fig. 1c) the forces along the shared axis may be neither of equal magnitude nor opposite sign.

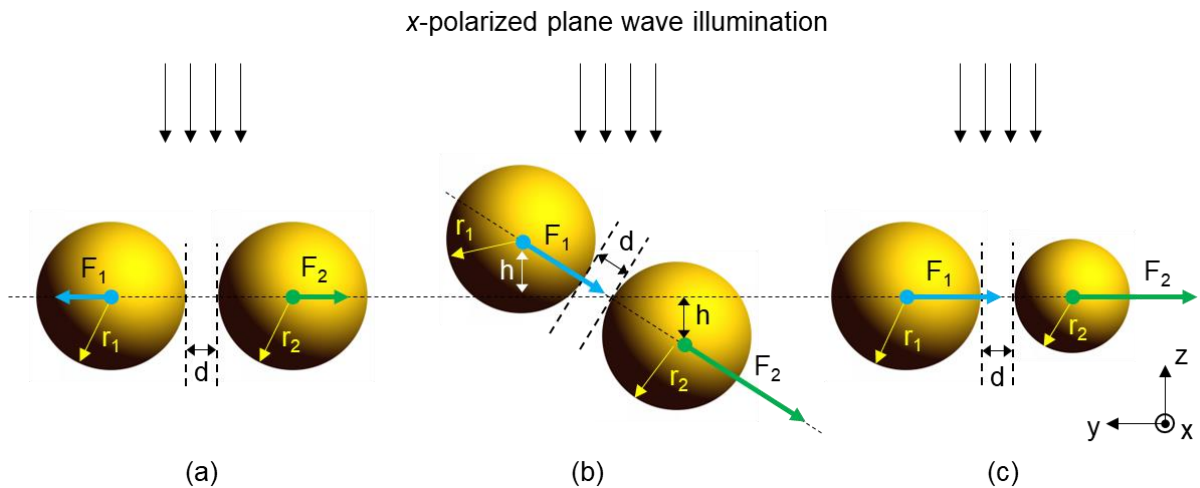

**Fig. 1. Light-induced nonreciprocal optical forces.** Optical forces $F_{1,2}$ acting on a pair of spherical gold nanoparticles along their centre-to-centre axis under illumination with x-polarized light at a wavelength of 550nm at an intensity of 1327 µW/µm² when the particles are: (a) of the same size and illuminated in phase; (b) of the same size and illuminated with a relative phase lag; (c) of different sizes. Parameter values are as follows: (a) $r_1$, $r_1$ = 60 nm; $d$ = 40 nm; $F_1$, $F_2$ = -1.5 fN. (b) $r_1$, $r_1$ = 60 nm; $d$ = 40 nm; $h$ = 40 nm; $F_1$ = 30 fN; $F_2$ = 64 fN. (c) $r_1$ = 60 nm; $r_2$ = 40 nm; $d$ = 40 nm; $F_1$ = 9 fN; $F_2$ = 5 fN.

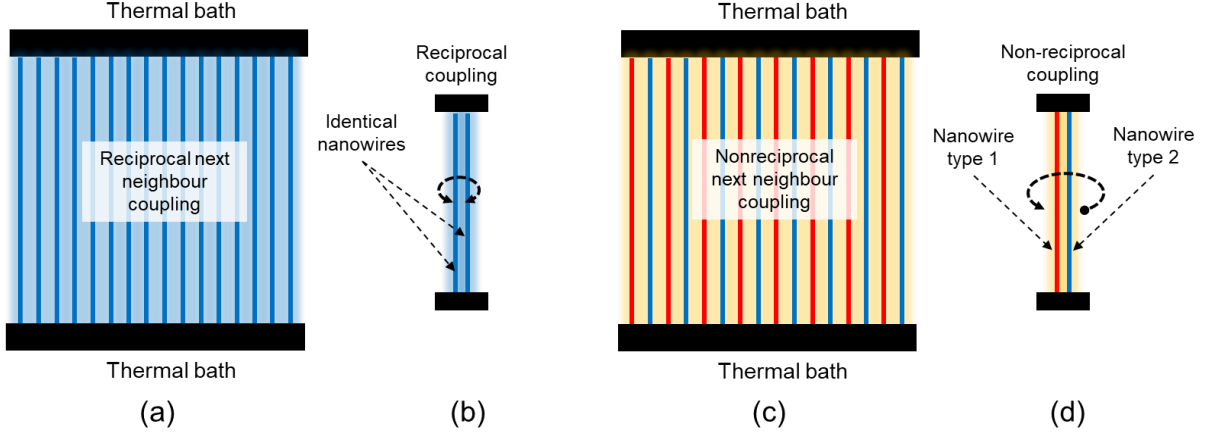

**Fig. 2. Schematics of nanowire metamaterials:** a) Metamaterial where all nanowires are identical. In the presence of light, they are reciprocally coupled; b) Isolated pair of nanowires with reciprocal coupling; c) Metamaterial consisting of alternating nanowires (color-coded) bearing different nanoparticles, which can be nonreciprocally coupled in the presence of light; d) Isolated pair of different nanowires with nonreciprocal coupling.

Let us now consider a simplified model of the classical continuous space-time crystal demonstrated in Ref. [8]: a two-dimensional array of nanowires decorated with plasmonic particles. We will examine two cases, where all nanowires in the array are decorated with the same type of plasmonic particles (Fig. 2a), and where there are two types of alternating nanowires decorated with different types of nanoparticles (Fig 2c). We assume that in the presence of light, neighbouring nanowires supporting identical nanoparticles interact reciprocally, while nanowires bearing different nanoparticles interact nonreciprocally.

We begin with an analysis of the motion of isolated pairs of nanowires with reciprocal (Fig. 2b) and nonreciprocal (Fig. 2d) interactions between them. The Brownian thermal motion of nanowires at non-zero temperatures is accounted for by assuming that they are connected to a common bath (frame of the metamaterial) at temperature $T$. Such a system is described by the Langevin model for linear oscillators with frequencies $\omega_{0i}$, masses $m_i$ and loss parameters $\gamma_i = \omega_{0i}/Q_i$:

$$\ddot{x}_i + \gamma_i \dot{x}_i + \omega_{0i}^2 x_i + \sum \xi_{ij}(x_i - x_j) = \sqrt{\frac{2k_B T \gamma_i}{m_i}} \eta_i(t) \qquad (1)$$

where $Q_i$ and $\eta_i(t)$ are quality factors and normalized white noise terms, and the parameter $\xi_{ij}$ describes light-induced coupling between oscillators.

We analyse the behaviour of such systems under various coupling conditions by numerically solving Eq. (1) using parameters of the oscillators $\omega_{0i}, m_i, \gamma_i$ close to those of the experimental system in Ref. [8]. From Maxwell tensor calculations, we have evaluated the electromagnetic forces acting between nanoparticles used in this work and found that they have a predominantly nonreciprocal nature. However, to illustrate the difference between reciprocal $\xi_{12} = \xi_{21} = \xi$ and nonreciprocal $\xi_{12} = -\xi_{21} = \xi^*$ coupling we consider both cases. We first study the case of two identical oscillators ($\omega_{01} = \omega_{02} = \omega_0$; $\gamma_1 = \gamma_2$). Figure 3a shows spectra of oscillator positions $x_1(t)$ and $x_2(t)$ in the absence of coupling between the oscillators, calculated from 1 ms segments of time series data: the spectra are identical and exhibit resonances at the oscillators' fundamental frequency $\omega_0$. If reciprocal coupling between oscillators is introduced (Fig. 2b) the resonances split in two: one remains at $\omega_0$, while a run-away resonance at $\widetilde{\omega}_0$ appears and is blue shifted ($\widetilde{\omega}_0 - \omega_0$ increasing) with increasing coupling coefficient $\xi$.

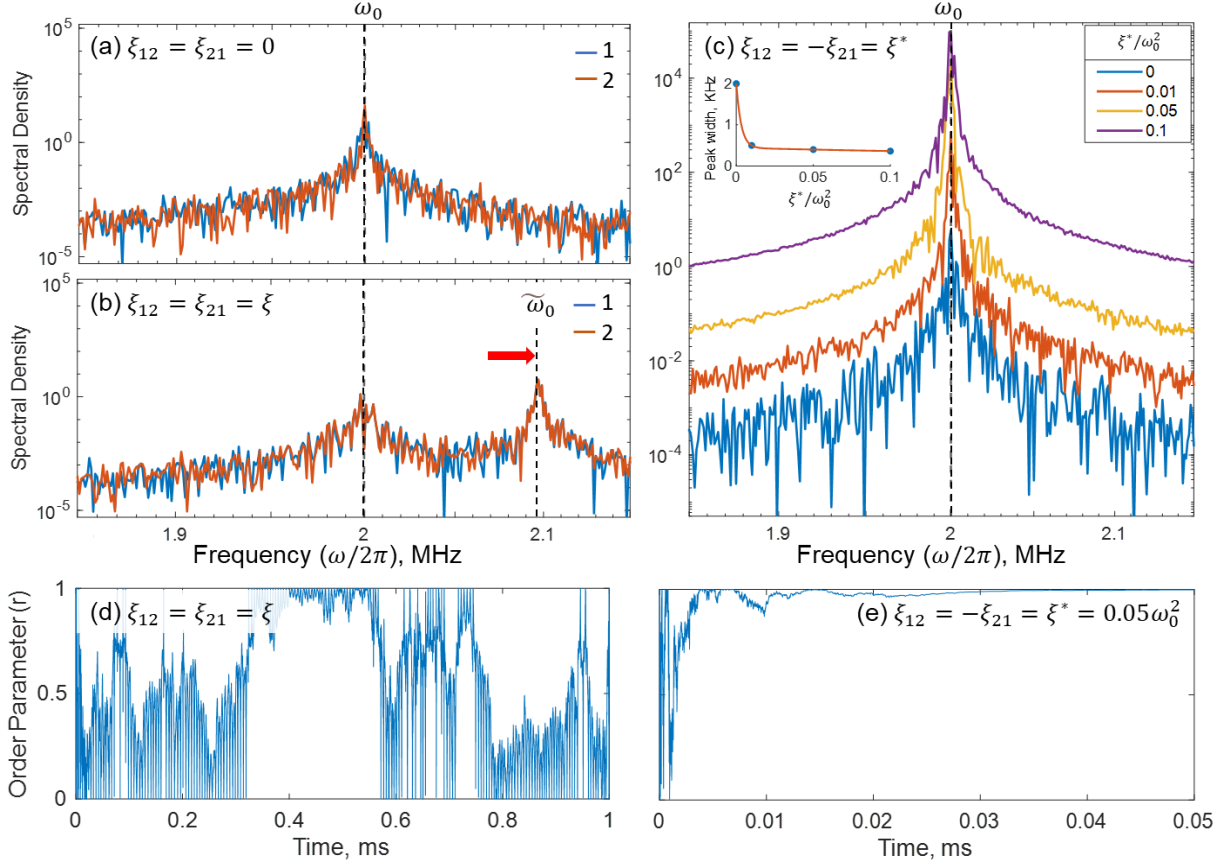

**Fig. 3. Dynamics of reciprocally coupled noisy oscillators.** (a, b) Spectral density of position for a pair of identical noise-driven oscillators [1 and 2] with natural frequency $\omega_0/2\pi = 2$ MHz, effective mass 1 pg and quality factor $Q = 1000$, at $T = 300$K in the case of: (a) no coupling between oscillators: $\xi_{12} = \xi_{21} = 0$; (b) reciprocal coupling: $\xi_{12} = \xi_{21} = \xi = 0.05\omega_0^2$; (c) Spectral density of position for one of two identical nonreciprocally coupled oscillators, for a selection of coupling strengths $\xi^*$ [as labelled]. The inset shows resonance width as a function of $\xi^*$. (d, e) Order parameter as a function of time for: (d) the reciprocally coupled system, from the moment in time when coupling reaches $\xi = 0.05\omega_0^2$; (e) nonreciprocal coupling from $\xi^* = 0.05\omega_0^2$ [the yellow trace in panel (c)].

Reversing the sign of $\xi$ reverses the direction of the spectral shift. Although the spectra of the oscillators' motion remain identical in the presence of reciprocal coupling, their movement is uncorrelated at any given value of $\xi$. This can be seen from the random and persistent variation of the order parameter $r(t) = \frac{1}{2} abs\{e^{i\varphi_1(t)} + e^{i\varphi_2(t)}\}$ as a function of time (Fig. 3d). Here the order parameter is a measure of the synchrony of the oscillators. Any stable value of $r$ means that the phase lag $\varphi_1(t) - \varphi_2(t)$ between oscillations is constant and they are synchronized; $r = 1$ corresponds to the oscillators moving in phase.

In the case of nonreciprocal coupling ($\xi_{12} = -\xi_{21} = \xi^*$), the oscillator dynamics are radically different (Fig. 3c): Here, we no longer plot separate spectra for the two oscillators, but rather just one of them. There is no resonance splitting in this case. Instead, the amplitude of oscillation grows with increasing $\xi^*$, and the resonance narrows (see inset to Fig. 3c). Moreover, relative noise decreases, and the synchronized oscillation resemble that of a single oscillator with amplitude and quality factor growing with $\xi^*$ (Fig 3c). Movements of the oscillators become correlated, and this happens faster for stronger nonreciprocal coupling. After the initial transition period, the order parameter asymptotically approaches $r = 1$, as illustrated in Fig. 3e.

We now consider synchronization due to nonreciprocal coupling between two oscillators with slightly different natural frequencies, $\omega_{01} = \omega_0 - \delta\omega$ and $\omega_{02} = \omega_0 + \delta\omega$, where $\gamma \ll \delta\omega \ll \omega_0$. This is a case of particular practical importance because manufacturing tolerances in real samples inevitably lead to a variation in the fundamental frequency of oscillation from nanowire to nanowire. In this case, synchronization is also achievable, but only above a threshold of nonreciprocal coupling strength $\xi_{th}^*$. Figure 4a shows the spectral density of relative position for a pair of oscillators for various values of nonreciprocal coupling strength. In the absence of coupling (blue line), two peaks are seen at the corresponding fundamental frequencies of oscillators. As $\xi^*$ approaches $\xi_{th}^* = 0.00504\omega_{0i}^2$, the peaks shift towards one another and collapse into a single resonance at $\omega_0$ (orange line). Further increasing the coupling strength leads to an abrupt, orders-of-magnitude exponential increase in oscillation amplitude and a decrease in relative noise, as illustrated by the yellow line. Synchronization of the oscillators' motion is also manifested in the order parameter $r$, which rapidly evolves (as a function of increasing coupling strength, Fig. 4b) to a stable value corresponding to a stable phase difference between the oscillators (which is a function of $\delta\omega$).

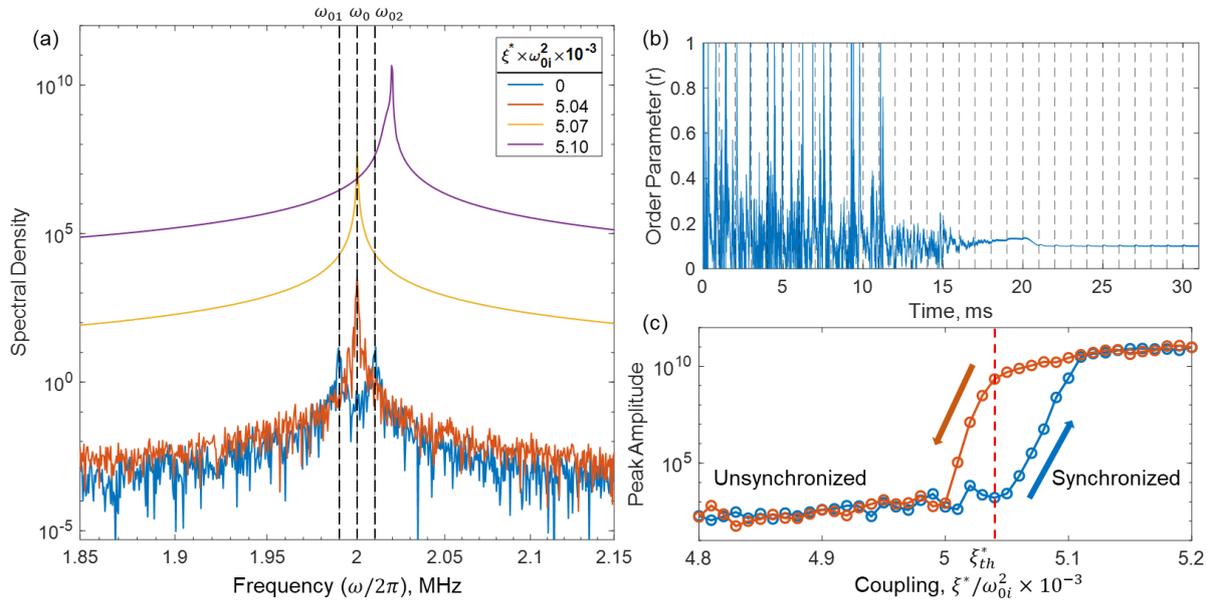

**Fig. 4. Dynamics of nonreciprocally coupled noisy oscillators.** (a) Spectral density of relative position for a pair of nonreciprocally coupled oscillators at various coupling strengths Here, the presence of nonlinearity is manifested by frequency pulling at high coupling strength, indicated by purple curve . The vertical dashed lines indicate the mean and fundamental frequencies of oscillators. (b) Order parameter as a function of time with coupling strength increasing from zero at time = 0 in increments of $0.00001\omega_{0i}^2$ at 1 ms intervals (as illustrated by the vertical dashed lines). (c) Peak amplitude of spectral density as a function of increasing (blue) and decreasing (orange) coupling strength. The vertical red line indicates the threshold coupling, $\xi_{th}^*$ for the onset of synchronization.

It should be noted here that the amplitude of synchronized oscillation of a pair of nonreciprocally coupled linear oscillators (above the $\xi_{th}^*$ threshold) grows with time. As such, short interval time series of $x_1(t)$ and $x_2(t)$ are used for calculating spectra. In Fig. 4, spectra are calculated for 2000 oscillation periods, i.e., over an interval of 1 ms during which oscillation amplitude grows by only a few percent. In real systems, increasing oscillation amplitude is constrained by the nonlinearity of the system, which in the case of nanowires anchored at both ends is known as the 'geometric nonlinearity'. This adds a term proportional to $x_i^3$ to the left-hand side of the equation of motion. The presence of this nonlinearity leads to saturation of the oscillation amplitude and to 'pulling' of the synchronized oscillation frequency away from $\omega_0$ towards higher frequencies, as illustrated by the purple line in Fig. 4a (where a term $10^9 \omega_{0i}^2 x_i^3$ is included in Eq. (1)). The saturation of oscillation amplitude as a function of increasing coupling strength is shown in Fig. 4c, which also reveals the hysteresis in the synchronization transition – an important characteristic of a first-order phase transition.

It should be emphasized here that the oscillators' nonlinearity is not needed for their synchronization, as it would be in the regimes of synchronization that depend on nonlinearity [14], it only moderates and stabilizes oscillation amplitudes.

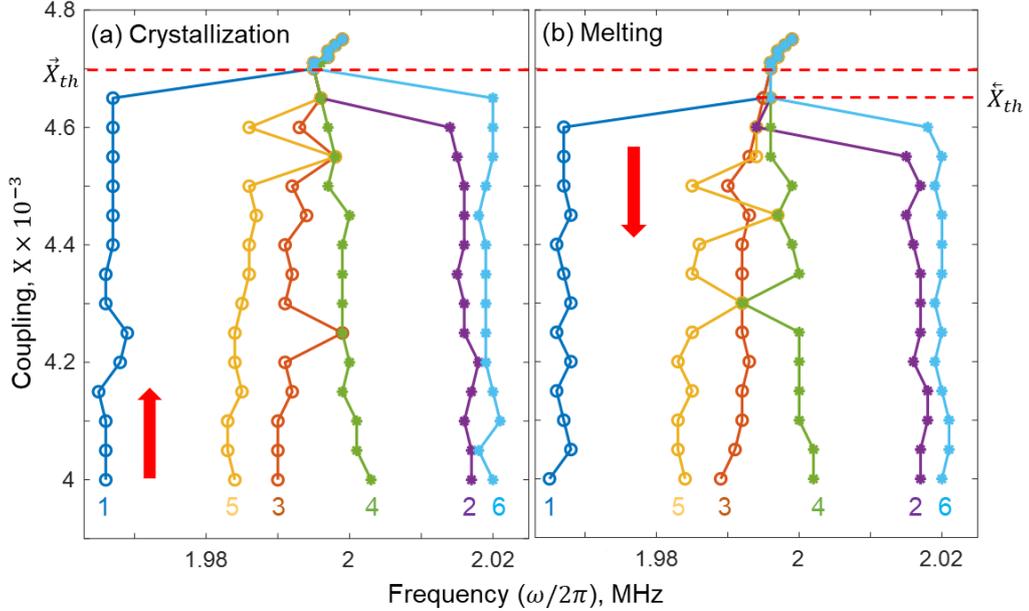

**Fig. 5. Crystallization and melting of a time crystal.** Synchronization (crystallization) and de-synchronization (melting) in an ensemble of six nonreciprocally coupled oscillators. The numerical annotations at the bottom of each line denote the position of the oscillator within the array.

Finally, we illustrate the synchronization process for an array of six inhomogeneously broadened oscillators each nonreciprocally coupled to their nearest neighbours. This provides for an illustration of one of the most characteristic features of the CTC regime – loss of ergodicity. We employ the same set of equations (1) as above, where indices $i, j$ may take values between 1 and 6. We assume that $\xi_{ij}/\omega_{0i}^2 = -\xi_{ji}/\omega_{0j}^2 = $ X for all values of $i, j$, and that oscillators are coupled to their near neighbours only. (There is no coupling between the first and the last oscillators in the array.) We also assume frequencies $\omega_{0,1-6}/2\pi =$ 1.961, 2.025, 1.979, 2.010, 1.974, 2.023 MHz and include the cubic nonlinearity term $10^9 \omega_{0i}^2 x_i^3$ in the equations of motion.

At low coupling strengths, the phase space of the system is rich, the oscillators move stochastically, and spectra contain six separate peaks at frequencies as illustrated at the bottom of Fig. 5 (where $X = 4.0$). With increasing coupling (Fig. 5a), some of the resonances merge and then demerge, temporarily creating 'domains' of neighbouring nanowires that oscillate in unison (e.g. $i = 3,4$ at $X = 4.25$). At $X = 4.65$ oscillators $i = 2,3,4,5$ become robustly synchronized and then at $\vec{X}_{th} = 4.7$ are joined by the remaining oscillators $i = 1,6$ (the right arrow denotes increasing coupling). Synchronization is complete: the oscillator ensemble's phase space has collapsed into a single unified trajectory, ergodicity of the system has been broken. With decreasing coupling the fully synchronized state breaks at $\overleftarrow{X}_{th} < \vec{X}_{th}$ (hysteresis). The restoration of stochastic motion also goes through the stage of disintegration and formation of domains as can be seen in Fig. 5b.

In conclusion, we have demonstrated that the recent observation of a light-induced continuous time crystal state in an array of interacting nanowires decorated with plasmonic nanoparticles can be comprehensively explained as a nonreciprocal phase transition, including spontaneous synchronization of the array and loss of its ergodicity.